\input harvmac
\noblackbox
 \def\p{\partial }
 \def\v{\tau}
   \def\cl{{\cal L}}
\lref\verl{
  E.~P.~Verlinde,
  ``On the Origin of Gravity and the Laws of Newton,''
  arXiv:1001.0785 [hep-th].
}
\lref\PadmanabhanVY{
  T.~Padmanabhan,
  ``Thermodynamical Aspects of Gravity: New insights,''
  Rept.\ Prog.\ Phys.\  {\bf 73}, 046901 (2010)
  [arXiv:0911.5004 [gr-qc]].
}

\lref\susskind{
  L.~Susskind,
  ``The World As A Hologram,''
  J.\ Math.\ Phys.\  {\bf 36}, 6377 (1995)
  [arXiv:hep-th/9409089].}
\lref\bousso{
  R.~Bousso,
  ``The holographic principle for general backgrounds,''
  Class.\ Quant.\ Grav.\  {\bf 17}, 997 (2000)
  [arXiv:hep-th/9911002].
  }
\lref\cremone{
  S.~Cremonini, K.~Hanaki, J.~T.~Liu and P.~Szepietowski,
  ``Higher derivative effects on eta/s at finite chemical potential,''
  Phys.\ Rev.\  D {\bf 80}, 025002 (2009)
  [arXiv:0903.3244 [hep-th]].
}
\lref\cremtwo{
  S.~Cremonini, K.~Hanaki, J.~T.~Liu and P.~Szepietowski,
  ``Black holes in five-dimensional gauged supergravity with higher
  derivatives,''
  JHEP {\bf 0912}, 045 (2009)
  [arXiv:0812.3572 [hep-th]].
}
\lref\BanerjeeZD{
  N.~Banerjee and S.~Dutta,
  ``Nonlinear Hydrodynamics from Flow of Retarded Green's Function,''
  arXiv:1005.2367 [hep-th].
}
\lref\PaulosYK{
  M.~F.~Paulos,
  ``Transport coefficients, membrane couplings and universality at
  extremality,''
  JHEP {\bf 1002}, 067 (2010)
  [arXiv:0910.4602 [hep-th]].
}

\lref\Polch{
  J.~Polchinski,
   ``Semi-holographic Fermi Liquids,''
  Talk at ``Strings 2010'' (2010).
}

\lref\gmss{
  R.~Gopakumar, S.~Minwalla, N.~Seiberg and A.~Strominger,
  ``OM Theory in Diverse Dimensions,''
  JHEP {\bf 0008}, 008 (2000)
  [arXiv:hep-th/0006062].
}

\lref\mark{
  M.~Van Raamsdonk,``Comments on quantum gravity and entanglement,''
  arXiv:0907.2939 [hep-th];
  ``Building up spacetime with quantum entanglement,''
  arXiv:1005.3035 [hep-th].
}
\lref\tadashi{
  T.~Nishioka, S.~Ryu and T.~Takayanagi,
  ``Holographic Entanglement Entropy: An Overview,''
  J.\ Phys.\ A  {\bf 42}, 504008 (2009)
  [arXiv:0905.0932 [hep-th]].
}

\lref\znaj{
R.~L.~Znajek
``The electric and magnetic conductivity of a Kerr hole'',
Mon.\ Not.\ R.\ Astron. Soc.\ {\bf 185}, 833 (1978).}
\lref\theo{
  T.~Jacobson,
  ``Thermodynamics of space-time: The Einstein equation of state,''
  Phys.\ Rev.\ Lett.\  {\bf 75}, 1260 (1995)
  [arXiv:gr-qc/9504004].
}
\lref\HartleZZ{
  J.~B.~Hartle,
  ``Tidal Friction in Slowly Rotating Black Holes,''
  Phys.\ Rev.\  D {\bf 8}, 1010 (1973).
}

\lref\HartleGY{
  J.~B.~Hartle,
  ``Tidal shapes and shifts on rotating black holes,''
  Phys.\ Rev.\  D {\bf 9}, 2749 (1974).
}


\lref\MorganPN{
  J.~Morgan, V.~Cardoso, A.~S.~Miranda, C.~Molina and V.~T.~Zanchin,
  ``Gravitational quasinormal modes of AdS black branes in d spacetime
  dimensions,''
  JHEP {\bf 0909}, 117 (2009)
  [arXiv:0907.5011 [hep-th]].
}


\lref\SonEM{
  D.~T.~Son and A.~O.~Starinets,
  ``Hydrodynamics of R-charged black holes,''
  JHEP {\bf 0603}, 052 (2006)
  [arXiv:hep-th/0601157].
}

\lref\MasDY{
  J.~Mas,
  ``Shear viscosity from R-charged AdS black holes,''
  JHEP {\bf 0603}, 016 (2006)
  [arXiv:hep-th/0601144].
}
\lref\MaedaBY{
  K.~Maeda, M.~Natsuume and T.~Okamura,
  ``Viscosity of gauge theory plasma with a chemical potential from  AdS/CFT,''
  Phys.\ Rev.\  D {\bf 73}, 066013 (2006)
  [arXiv:hep-th/0602010].
}
\lref\GeAK{
  X.~H.~Ge, Y.~Matsuo, F.~W.~Shu, S.~J.~Sin and T.~Tsukioka,
  ``Density Dependence of Transport Coefficients from Holographic
  Hydrodynamics,''
  Prog.\ Theor.\ Phys.\  {\bf 120}, 833 (2008)
  [arXiv:0806.4460 [hep-th]].
}

\lref\MyersIJ{
  R.~C.~Myers, M.~F.~Paulos and A.~Sinha,
  ``Holographic Hydrodynamics with a Chemical Potential,''
  JHEP {\bf 0906}, 006 (2009)
  [arXiv:0903.2834 [hep-th]].
}


\lref\ChircoXX{
  G.~Chirco, C.~Eling and S.~Liberati,
  ``The universal viscosity to entropy density ratio from entanglement,''
  arXiv:1005.0475 [hep-th].
}

\lref\IqbalBY{
  N.~Iqbal and H.~Liu,
  ``Universality of the hydrodynamic limit in AdS/CFT and the membrane
  paradigm,''
  Phys.\ Rev.\  D {\bf 79}, 025023 (2009)
  [arXiv:0809.3808 [hep-th]].
}

\lref\PaulosYK{
  M.~F.~Paulos,
  ``Transport coefficients, membrane couplings and universality at
  extremality,''
  JHEP {\bf 1002}, 067 (2010)
  [arXiv:0910.4602 [hep-th]].
}

\lref\BhattacharyyaMZ{
  S.~Bhattacharyya, R.~Loganayagam, I.~Mandal, S.~Minwalla and A.~Sharma,
  ``Conformal Nonlinear Fluid Dynamics from Gravity in Arbitrary Dimensions,''
  JHEP {\bf 0812}, 116 (2008)
  [arXiv:0809.4272 [hep-th]].
}

\lref\DamourCG{
  T.~Damour,
  ``Black Hole Eddy Currents,''
  Phys.\ Rev.\  D {\bf 18}, 3598 (1978).
}
\lref\oldDamour{
   T.~Damour,
  ``Surface effects in Black Hole Physics,'' in {\it Proceedings of the second Marcel Grossmann Meeting on general Relativity},
  Ed. R.Ruffini, North-Holland, 1982. }

\lref\thesDamour{
   T.~Damour, in:
   ``Quelques propri\'et\'es m\'ecaniques, \'electromagn\'etiques, thermodynamiques et quantiques des trous noirs''; Th\`ese de Doctorat d'Etat, Universit\'e
   Pierre et Marie Curie, Paris VI, 1979.}

\lref\HawkingHY{
  S.~W.~Hawking and J.~B.~Hartle,
  ``Energy And Angular Momentum Flow Into A Black Hole,''
  Commun.\ Math.\ Phys.\  {\bf 27}, 283 (1972).
}

\lref\GourgoulhonCH{
  E.~Gourgoulhon,
  ``A generalized Damour-Navier-Stokes equation applied to trapping horizons,''
  Phys.\ Rev.\  D {\bf 72}, 104007 (2005)
  [arXiv:gr-qc/0508003].
}
\lref\McGreevyXE{
  J.~McGreevy,
  ``Holographic duality with a view toward many-body physics,''
  [arXiv:0909.0518 [hep-th]].
}

\lref\SachdevCH{
  S.~Sachdev,
  ``Condensed matter and AdS/CFT,''
  arXiv:1002.2947 [hep-th].
}

\lref\SusskindJS{
  L.~Susskind and J.~Lindesay,
  ``An introduction to black holes, information and the string theory
  revolution: The holographic universe,''
{\it  Hackensack, USA: World Scientific (2005) 183 p.}
}

\lref\ThorneIY{
  K.~S.~.~Thorne, R.~H.~.~Price and D.~A.~.~Macdonald,
  ``Black Holes: the Membrane Paradigm,''
{\it  New Haven, USA: Yale Univ. Pr.  (1986) 367p.}
}

\lref\PriceYY{
  R.~H.~Price and K.~S.~Thorne,
  ``Membrane viewpoint on black holes: properties and evolution of the stretched horizon,''
  Phys.\ Rev.\  D {\bf 33}, 915 (1986).
}
\lref\DamourJI{
  T.~Damour and M.~Lilley,
  ``String theory, gravity and experiment,''
  arXiv:0802.4169 [hep-th].
}



\lref\kss{}

\lref\star{
  A.~O.~Starinets,
  ``Quasinormal spectrum and the black hole membrane paradigm,''
  Phys.\ Lett.\  B {\bf 670}, 442 (2009)
  [arXiv:0806.3797 [hep-th]].
}
\lref\SonVK{
  D.~T.~Son and A.~O.~Starinets,
  ``Viscosity, Black Holes, and Quantum Field Theory,''
  Ann.\ Rev.\ Nucl.\ Part.\ Sci.\  {\bf 57}, 95 (2007)
  [arXiv:0704.0240 [hep-th]].
}
\lref\ks{
  P.~K.~Kovtun and A.~O.~Starinets,
  ``Quasinormal modes and holography,''
  Phys.\ Rev.\  D {\bf 72}, 086009 (2005)
  [arXiv:hep-th/0506184].
}
\lref\KovtunWP{
  P.~Kovtun, D.~T.~Son and A.~O.~Starinets,
  ``Holography and hydrodynamics: Diffusion on stretched horizons,''
  JHEP {\bf 0310}, 064 (2003)
  [arXiv:hep-th/0309213].
}
\lref\PolicastroTN{
  G.~Policastro, D.~T.~Son and A.~O.~Starinets,
  ``From AdS/CFT correspondence to hydrodynamics. II: Sound waves,''
  JHEP {\bf 0212}, 054 (2002)
  [arXiv:hep-th/0210220].
}
\lref\KodamaFA{
  H.~Kodama, A.~Ishibashi and O.~Seto,
  ``Brane world cosmology: Gauge-invariant formalism for perturbation,''
  Phys.\ Rev.\  D {\bf 62}, 064022 (2000)
  [arXiv:hep-th/0004160].
}
\lref\PSS{
  G.~Policastro, D.~T.~Son and A.~O.~Starinets,``The shear viscosity of strongly coupled N = 4 supersymmetric Yang-Mills
  plasma,''
  Phys.\ Rev.\ Lett.\  {\bf 87}, 081601 (2001)
  [arXiv:hep-th/0104066];
``From AdS/CFT correspondence to hydrodynamics,''
  JHEP {\bf 0209}, 043 (2002)
  [arXiv:hep-th/0205052].
}
\lref\BuchelSK{
  A.~Buchel, J.~Escobedo, R.~C.~Myers, M.~F.~Paulos, A.~Sinha and M.~Smolkin,
  ``Holographic GB gravity in arbitrary dimensions,''
  JHEP {\bf 1003}, 111 (2010)
  [arXiv:0911.4257 [hep-th]].
}

\lref\BuchelVZ{
  A.~Buchel, R.~C.~Myers and A.~Sinha,
  ``Beyond eta/s = 1/4pi,''
  JHEP {\bf 0903}, 084 (2009)
  [arXiv:0812.2521 [hep-th]].
}
\lref\BriganteNU{
  M.~Brigante, H.~Liu, R.~C.~Myers, S.~Shenker and S.~Yaida,
  ``Viscosity Bound Violation in Higher Derivative Gravity,''
  Phys.\ Rev.\  D {\bf 77}, 126006 (2008)
  [arXiv:0712.0805 [hep-th]].
}

\lref\KatsMQ{
  Y.~Kats and P.~Petrov,
  ``Effect of curvature squared corrections in AdS on the viscosity of the dual
  gauge theory,''
  JHEP {\bf 0901}, 044 (2009)
  [arXiv:0712.0743 [hep-th]].
}
\lref\MyersYI{
  R.~C.~Myers, M.~F.~Paulos and A.~Sinha,
  ``Quantum corrections to eta/s,''
  Phys.\ Rev.\  D {\bf 79}, 041901 (2009)
  [arXiv:0806.2156 [hep-th]].
}

\lref\KlebanovHB{
  I.~R.~Klebanov and M.~J.~Strassler,
  ``Supergravity and a confining gauge theory: Duality cascades and
  chiSB-resolution of naked singularities''
  JHEP {\bf 0008}, 052 (2000)
  [arXiv:hep-th/0007191].
}
\lref\HorowitzCD{
  G.~T.~Horowitz and A.~Strominger,
  ``Black strings and P-branes''
  Nucl.\ Phys.\  B {\bf 360}, 197 (1991).
}
\lref\GiddingsMI{
  S.~B.~Giddings and A.~Strominger,
  Phys.\ Rev.\ Lett.\  {\bf 67}, 2930 (1991).
}

\lref\BuchelHW{
  A.~Buchel,
  ``N = 2* hydrodynamics,''
  Nucl.\ Phys.\  B {\bf 708}, 451 (2005)
  [arXiv:hep-th/0406200].
}

\lref\BenincasaFU{
  P.~Benincasa, A.~Buchel and R.~Naryshkin,
  ``The shear viscosity of gauge theory plasma with chemical potentials,''
  Phys.\ Lett.\  B {\bf 645}, 309 (2007)
  [arXiv:hep-th/0610145].
}

\lref\HerzogFN{
  C.~P.~Herzog,
  ``The hydrodynamics of M-theory,''
  JHEP {\bf 0212}, 026 (2002)
  [arXiv:hep-th/0210126].
}
\lref\HerzogKE{
  C.~P.~Herzog,
  ``The sound of M-theory,''
  Phys.\ Rev.\  D {\bf 68}, 024013 (2003)
  [arXiv:hep-th/0302086].
}
\lref\ElingPB{
  C.~Eling, I.~Fouxon and Y.~Oz,
  ``The Incompressible Navier-Stokes Equations From Membrane Dynamics,''
  Phys.\ Lett.\  B {\bf 680}, 496 (2009)
  [arXiv:0905.3638 [hep-th]].
}
\lref\ElingSJ{
  C.~Eling and Y.~Oz,
  ``Relativistic CFT Hydrodynamics from the Membrane Paradigm,''
  JHEP {\bf 1002}, 069 (2010)
  [arXiv:0906.4999 [hep-th]].
}

\lref\HartnollKX{
  S.~A.~Hartnoll, C.~P.~Herzog and G.~T.~Horowitz,
  ``Holographic Superconductors,''
  JHEP {\bf 0812}, 015 (2008)
  [arXiv:0810.1563 [hep-th]].
}

\lref\HartnollSZ{
  S.~A.~Hartnoll,
  ``Lectures on holographic methods for condensed matter physics,''
  Class.\ Quant.\ Grav.\  {\bf 26}, 224002 (2009)
  [arXiv:0903.3246 [hep-th]].
}

\lref\KovtunKX{
  P.~Kovtun and A.~Ritz,
  ``Universal conductivity and central charges,''
  Phys.\ Rev.\  D {\bf 78}, 066009 (2008)
  [arXiv:0806.0110 [hep-th]].
}

\lref\BuchelTZ{
  A.~Buchel and J.~T.~Liu,
  ``Universality of the shear viscosity in supergravity,''
  Phys.\ Rev.\ Lett.\  {\bf 93}, 090602 (2004)
  [arXiv:hep-th/0311175].
}


\lref\BhattacharyyaKQ{
  S.~Bhattacharyya, S.~Minwalla and S.~R.~Wadia,
  ``The Incompressible Non-Relativistic Navier-Stokes Equation from Gravity,''
  JHEP {\bf 0908}, 059 (2009)
  [arXiv:0810.1545 [hep-th]].
}

\lref\BhattacharyyaJI{
  S.~Bhattacharyya, R.~Loganayagam, S.~Minwalla, S.~Nampuri,
S.~P.~Trivedi and S.~R.~Wadia,
  ``Forced Fluid Dynamics from Gravity,''
  JHEP {\bf 0902}, 018 (2009)
  [arXiv:0806.0006 [hep-th]].
}

\lref\ZamolodchikovGT{
  A.~B.~Zamolodchikov,
  ``Irreversibility of the Flux of the Renormalization Group in a 2D Field
  Theory,''
  JETP Lett.\  {\bf 43}, 730 (1986)
  [Pisma Zh.\ Eksp.\ Teor.\ Fiz.\  {\bf 43}, 565 (1986)].
}

\lref\FreedmanGP{
  D.~Z.~Freedman, S.~S.~Gubser, K.~Pilch and N.~P.~Warner,
  ``Renormalization group flows from holography supersymmetry and a
  c-theorem,''
  Adv.\ Theor.\ Math.\ Phys.\  {\bf 3}, 363 (1999)
  [arXiv:hep-th/9904017].
}
\lref\ascv{
  A.~Strominger and C.~Vafa,
  ``Microscopic Origin of the Bekenstein-Hawking Entropy,''
  Phys.\ Lett.\  B {\bf 379}, 99 (1996)
  [arXiv:hep-th/9601029].
}
\lref\MaldacenaRE{
  J.~M.~Maldacena,
  ``The large N limit of superconformal field theories and supergravity,''
  Adv.\ Theor.\ Math.\ Phys.\  {\bf 2}, 231 (1998)
  [Int.\ J.\ Theor.\ Phys.\  {\bf 38}, 1113 (1999)]
  [arXiv:hep-th/9711200].
}

\Title{\vbox{\baselineskip12pt
}}
{\vbox{\centerline {Wilsonian Approach to }
\centerline {Fluid/Gravity Duality} }}
\centerline{ Irene Bredberg, Cynthia Keeler,  Vyacheslav Lysov and Andrew Strominger}

\bigskip\centerline{\it Center for the Fundamental Laws of Nature}
\centerline{\it Harvard University, Cambridge, MA 02138 USA }

\vskip .3in
\centerline{\bf Abstract}
  The problem of gravitational fluctuations confined inside a finite cutoff at radius $r=r_c$ outside the horizon in a general class of black hole geometries is considered. Consistent boundary conditions at both the cutoff surface and the horizon are found and the resulting modes analyzed.
  For general cutoff $r_c$  the dispersion relation is shown at  long wavelengths to be that of a linearized Navier-Stokes fluid living on the cutoff surface. A cutoff-dependent line-integral formula for the diffusion constant $D(r_c)$ is derived. The dependence on $r_c$ is interpreted as renormalization group (RG) flow in the fluid.
 Taking the cutoff to infinity in an asymptotically AdS context, the formula for $D(\infty)$ reproduces as a special case well-known results derived using  AdS/CFT. Taking the cutoff to the horizon, the effective speed of sound goes to infinity, the fluid becomes incompressible and the Navier-Stokes dispersion relation becomes exact. The resulting universal formula for the diffusion constant $D(horizon)$ reproduces old results from the membrane paradigm.
 Hence the old membrane paradigm results and new AdS/CFT results are related by RG flow.
 RG flow-invariance of the viscosity to entropy ratio $\eta \over s$ is shown to follow from the first law of thermodynamics together with isentropy of radial evolution in classical gravity.  The ratio is expected to run when quantum gravitational corrections are included.  \smallskip

\Date{}
\listtoc
\writetoc

\newsec{Introduction}

If you disturb a black hole, the horizon responds like a viscous fluid \refs{\HawkingHY,\HartleZZ,\HartleGY}. In particular, damped shear waves propagate outward from the disturbance.
The fluid viscosity  was computed to be\foot{   Throughout this paper we use $G$ for Newton's constant  and set $\hbar=k_B=1$.}  $\eta={1 \over 16\pi G}$  in a prescient 1978 thesis by Damour \thesDamour, see also \refs{\DamourCG\znaj\oldDamour
\PriceYY \ThorneIY- \DamourJI}.
Dividing this viscosity by the Bekenstein-Hawking horizon entropy density $s={1 \over 4G}$ yields the dimensionless ratio ${\eta \over s}={1\over 4 \pi}$. The ingredients of the computation are only
 the basic equations of general relativity and (in the computation of $s$) quantum mechanics.

A quarter century later, ${\eta \over  s}={1 \over 4 \pi}$ reappeared in a  beautiful paper by Policastro, Starinets and Son (PSS)\PSS, see also  \refs{\KovtunWP\KodamaFA \HerzogFN \PolicastroTN  \BuchelHW  \ks  \BenincasaFU \KatsMQ  \BriganteNU \KovtunKX \star\BhattacharyyaKQ \BuchelVZ \BuchelSK \cremtwo\MyersIJ -\cremone} and the review {\SonVK }. They considered the hydrodynamics of supersymmetric ${\cal N}=4$ $U(N)$ gauge theory. Stress tensor correlators at spatial infinity were computed in an AdS-black-brane geometry giving  $\eta={\pi \over 8}N^2T^3$ and $s={\pi^2\over 2} N^2 T^3$. This again yields ${\eta \over s}={1 \over 4 \pi}$. The ingredients of this computation are much more extensive than that of Damour and include string theory, supersymmetric gauge theories and AdS/CFT.

On the face of it, the Damour and PSS computations are very different. However, there are strong indications that they are related. Firstly, they both relate a theory of gravity to a ``dual'' fluid theory living in one fewer dimension; without the radial direction. Secondly, both approaches lead to the same numerical ratio for ${\eta \over s}$. An important difference is that the Damour calculation is performed at the black hole horizon $r=r_h$,
while the PSS calculation is performed at spatial infinity $r=\infty$. Both the basic relation between redshift and radius\foot{Radial transformations are referred to as renormalization already in \PriceYY.} and, in the special context of string theory, results from AdS/CFT suggest that from the point of view of the fluid theory (which does not have an $r$ coordinate) changing $r$ is equivalent to renormalization group (RG) flow. Hence one expects the Damour calculation to be related to the PSS calculation  by some kind of RG flow into the IR. This view is advocated in  \refs{\KovtunWP, \BuchelTZ,\SonVK, \IqbalBY\ElingPB\ElingSJ\PaulosYK-\BanerjeeZD}.\foot{These references typically study RG flow by looking at the $r$-dependence of correlators  whose boundary conditions are imposed in an asymptotically AdS region. This corresponds to choosing a specific UV completion. In this paper we will formulate
the problem in a way that does not involve such a choice.}  In order to verify this expectation, one must first define what one means by the gravity and  fluid theory associated to finite $r$. In this paper, among other things,  we propose  a precise definition
of the finite $r$ theory and show that the expectation is indeed realized.

The basic idea is to introduce a cutoff surface $\Sigma_c$ at some fixed radius $r=r_c$
outside the black hole or brane.  We then impose ingoing  boundary conditions at the horizon and fix the induced metric on $\Sigma_c$.  These boundary conditions do not fully specify a solution. The problem is then to identify the remaining internal degrees of freedom and describe their dynamics.  We solve this to linear order in the internal fluctuations (in appropriate expansions) and show they correspond to those of a fluid.  A formula is derived for the diffusion rate and other hydrodynamic quantities, which generically run as a function of the cutoff $r_c$. We hope that it is possible to extend our approach beyond the next-to-leading order considered here, but we defer that problem to future work.

This reformulation of fluid/gravity duality is the analog - or holographic dual - of Wilson's reformulation of quantum field theory.  Wilson did not insist on an ultraviolet completion of quantum field theory, and we do not insist on an asymptotically AdS region of the geometry.   Specifying
the couplings at the Wilsonian cutoff  $\Lambda_W$ is the analog of specifying the boundary conditions for the induced metric (and other fields if present) on $\Sigma_c$ .  If we scatter fields at energies below $\Lambda_W$, we needn't know anything about the theory at energies above $\Lambda_W$. Similarly if we disturb a black hole by throwing something at it from the radius $r_c$, we needn't know anything  about the geometry outside $r_c$.\foot{In this analogy, large $r_c$ corresponds to large $\Lambda_W$, but a precise functional relation will not be found herein.} One advantage of the Wilsonian  approach is that  a much broader class of theories can be discussed.

In addition to providing new methods of computation and broadening the space of applications, several new qualitative insights are gained in this approach.  As the cutoff is taken to the horizon ($r_c \to r_h$), the relevant geometry is simply Rindler space, and the transport coefficients all approach simply computable universal values which are largely insensitive to details of the geometry or matter couplings.\foot{This remains true as long as the linearized gravity fluctuations are governed by the linearized Einstein equation, which is not the case
with higher derivative corrections.} In particular ${\eta \over s}\to {1 \over 4\pi}$.
Moreover the fluctuations dissipate according to the linearized Navier-Stokes equations, with $no ~higher$-$derivative~corrections$.  Hence our near horizon scaling is the geometric version of the low-velocity scaling in which fluids are governed by (as it turns out incompressible) Navier-Stokes. At any finite $r_c>r_h$ there are infinitely many higher derivative corrections   to the Navier-Stokes dispersion relation, and one computes only the leading term at long wavelengths. The leading dispersion constant in general runs and does not take a universal value at radial infinity. If we specialize to an asymptotically  AdS black brane and take $r_c \to \infty$, our computations
are all in manifest agreement with the usual AdS/CFT definitions of the transport coefficients.  All of this supports the picture that Damour was computing in the IR of the dual fluid theory while PSS were computing in the UV. The extra ingredients required for the PSS calculation are the extra ingredients needed to specify a theory all the way up to the UV, while the universality of $\eta \over s$ is a characteristic of the IR fixed point and needs only the IR near-horizon Rindler space.

Interestingly enough, in contrast to the generic transport coefficient, in the classical gravity limit the particular ratio $\eta \over s$ typically does $not$ run and equals ${1 \over 4\pi}$ everywhere. This is why the UV and IR PSS and Damour calculations agree for this quantity. Although still partially mysterious to us, we show that this RG-flow invariance stems from the first law of thermodynamics applied to the radial flow, together with the fact that, in the classical gravity limit, there is no entropy except horizon entropy and the flow is therefore isentropic. This will not be the case when quantum corrections are included, as there is then entropy in the gas of Hawking radiation as well as entanglement entropy across $\Sigma_c$. Therefore we expect $\eta \over s$ to run at the quantum level.

\subsec{Applications}
  This work has several distinct applications and motivations.

\noindent{$\bullet$ \it Condensed matter / cold atoms}

  One possible application is to the recent exciting work (for reviews see \refs{\HartnollSZ, \McGreevyXE, \SachdevCH}) on trying to solve or understand a variety of condensed matter and cold atomic systems  via the construction
of holographic gravity duals.  Most of this has involved asymptotically AdS geometries.  If this program is to succeed, it cannot depend on the existence of such an AdS region (or underlying string theory), as one thing we know for sure about
real world materials is that they are not conformally invariant in the UV, and certainly do not become supersymmetric $N=4$ Yang-Mills or critical  string theory!  Rather success of this program should depend on the more universal features of the holographic relation between gravity and matter systems. The existence of such a universal relation became  clear already in the 70's, but many aspects of its form and origin remain(ed) deeply mysterious. The emergence of partially soluble examples within string theory has enabled us to guess much about how this more general relation must work. In this paper, we have set up a framework which attempts to liberate the essential features of fluid/gravity duality  from the confines of the string theory/AdS context. We hope that this will enable a broader class of applications, and circumvent the often unnecessary discussions of the asymptotic AdS region.\foot{This view has already begun to emerge in various discussions. For example, in a recent talk {\refs \Polch}, Polchinski
drew a line at the inner (IR) boundary of the AdS region and threw the AdS region out because it was irrelevant to the IR physics under consideration.  In this paper, we attempt to understand how the boundary conditions at Polchinski's  line (our $\Sigma_C$) are defined so that we need never draw the AdS region in the first place.}

\noindent{$\bullet$ \it The Navier-Stokes equation}

The Einstein and Navier-Stokes equation are two of the most important differential equations in physics and mathematics. Any direct relation between them is extremely interesting.  For example, it was suggested in \refs{\BhattacharyyaJI} that the problem of Navier-Stokes turbulence might be mapped to a problem in general relativity, with the different scales appearing in turbulent phenomena being mapped to different radii in the dual geometry.    We hope that our approach will facilitate this mapping, as one need not include an AdS region (irrelevant for turbulence), and the forcing term can be supplied  by time-dependent boundary conditions on $\Sigma_c$.  Of course, for this perspective to be useful, we must understand the duality beyond the linearized fluctuations considered in this paper.

\noindent{$\bullet$\it The gravity/thermo/quantum triangle }

The previous two applications aim to employ gravity to solve problems in other areas of physics or math. However it is possible that the ideas involved  have more than technical utility.
Ever since the seminal work of Bekenstein and Hawking, it has been clear that there is
a deep and fundamental relation between gravity, thermodynamics and quantum mechanics, while its detailed form and origin was and is largely mysterious. Much light has been shed on this triangle  in the specific context of string theory
following the microscopic description of black holes as a finite temperature two dimensional CFT \ascv\ and its higher-dimensional generalization to AdS/CFT \MaldacenaRE.
It seems likely  that the basic triangular relationships transcend string theory and AdS/CFT, although lessons from string theory are likely useful guides for unraveling the more general picture.

It is our hope that the attempt here to generalize fluid/gravity duality away from the stringy context to its most essential ingredients may be useful in understanding this triangle. Much of the current work on fluid/gravity duality attempts to learn about matter systems from gravity: we would like to  reverse the arrow towards an  understanding of quantum gravity.  In this regard, there may be interesting connections
with previous work in this direction including in particular \refs{ \theo, \verl, \mark}.
\subsec{Outline}

This paper is organized as follows. In section 2 we write down the most general $p+2$-dimensional general geometry which, on symmetry grounds, could serve as a holographic dual to a
fluid in $p+1$ dimensions.  Explicit expressions are given for the asymptotically flat and asymptotically AdS black branes with and without charges to serve as illustrative examples.

Section 3 treats the case of an electromagnetic field propagating on these geometries as a simple warm up. Dirichlet (ingoing) boundary conditions are imposed at the cutoff surface $r=r_c$ (horizon $r=r_h$). It is then shown in a long-wavelength expansion that the remaining dynamical modes are described by a charge density which evolves according to Fick's law in $p+1$ flat dimensions. The diffusion constant is given by a line integral of certain metric coefficients from $r_h$ to $r_c$, and ``runs'' as $r_c$ is varied.  If the cutoff $r_c$ is taken to $r_h$, no long wavelength expansion is needed and the Fick law becomes exact. Moreover it is shown in this limit that, after carefully normalizing by the divergent local Unruh temperature, the diffusion constant approaches a universal constant determined by the properties of Rindler space.

In section 4 the analysis is adapted to linearized gravity fluctuations.  After fixing the cutoff and horizon boundary conditions, the vector or shear modes are shown to obey the linearized Navier-Stokes equation in a long wavelength expansion, and a running formula for the diffusion constant is derived. As the cutoff is taken to the horizon, the linearized Navier-Stokes equation becomes exact and the constant is shown to approach the universal Damour value. Some special features of the RG flow for the gravitational case are also discussed. Tensor modes are shown to have no dynamics in the appropriate limit, while there is a dynamical ``sound'' mode in the scalar sector. Very interestingly,  the effective speed of sound goes to infinity and hence the sound  mode decouples as the cutoff is taken to the horizon. This means the fluid is becoming incompressible.  Specific examples of the charged and neutral  AdS black branes and asymptotically flat $S^3$-reduced NS5 branes are worked out in detail.

In section 5 we introduce the Brown-York stress tensor $t_{ab}$ on the cutoff surface.
Prior to this point only equations of motion have been used so entropy, viscosity, energy and pressure (which depend on the normalization of the action) could not be discussed. We show that $t_{ab}$ not only is conserved with our boundary conditions but takes the form of a fluid stress tensor (to linear order). We compute the thermodynamic quantities in terms of the spacetime geometry.

In section 6 we compute  the viscosity to entropy ratio $\eta \over s$ and show that, under rather general assumptions, the radial RG evolution equations imply it is cutoff independent and equal to $1 \over 4 \pi$ for Einstein gravity. It is shown that these radial equations are nothing but - in the fluid picture - the first law of thermodynamics for isentropic variations in disguise. The radial flow is isentropic because in classical gravity there is no entropy outside the black hole. It is accordingly suggested that the RG-invariance of $\eta \over s$ will be violated by quantum gravity corrections.

\newsec{Background geometry}
\subsec{The general case}
   In this paper we are interested in studying the dynamics of fluids in $p$ flat space dimensions and one flat time dimension. The holographic dual of such a fluid in its ground state should be a $p+2$-dimensional spacetime geometry, with isometries generating the Euclidean group of $p$-dimensional rotations/translations plus time translations. The corresponding  line element can be written in the form
 \eqn\dsa{ds_{p+2}^2=-h(r)d\v ^2+2d\v dr+e^{2t(r)}dx^idx_i,}
 where the index $i=1,...p$ here and hereafter is raised and lowered with $\delta_{ij}$.
 Lines of constant $\v $ and varying $r$ are null.  We consider the case where there is a horizon $r_h$ at which
 \eqn\dsaz{h(r_h)=0,}
 and $h(r)$ is positive for $r>r_h$.
 Lines of constant $r=r_h$ and varying $\v $ are the  null generators of the future horizon, while
those of constant $r>r_h$ and varying $\v $ are timelike and accelerated. For convenience we  choose the scaling of the spatial $x^i$ coordinate so that
\eqn\jja{t(r_h)=0.}

A special role will be played by the ``cutoff'' surfaces $\Sigma_c$ of constant $r=r_c>r_h$.
The induced metric on such a surface is flat $p+1$-dimensional Minkowski space
    \eqn\ykj{ds_{p+1}^2=-h\left(d\v -{dr \over h}\right)^2+e^{2t}dx^idx_i,}
    with $\v $ the time coordinate.
    We will sometimes collectively denote the Minkowskian coordinates by
    \eqn\vft{x^a\sim (x^i,\v ),~~~~a=0,\cdots p.} It is convenient to
introduce proper intrinsic coordinates on $\Sigma_c$
    \eqn\dio{x^0_c=\sqrt{h(r_c)}\v ,~~~x^i_c=e^{t(r_c)}x^i.} The advantage of  these coordinates is that the induced metric is simply
   \eqn\ddkn{ds_{p+1}^2=\eta_{ab}dx^a_cdx^b_c,}
   so that they directly measure proper distances on $\Sigma_c$.
Full bulk coordinates will be denoted
\eqn\vft{x^\mu \sim (x^i,\v ,r),~~~~\mu=0,\cdots p+1.}
 We denote  by $\ell^\mu$ the normal satisfying
    \eqn\dxw{\ell^\mu \p_\mu =(\p_\v +h\p_r),~~~~\ell^2=h.}
At $r=r_h$, $\ell$ is null, normal and tangent to the future horizon.
\subsec{Some special cases}
Here we collect some specific examples which will be used as illustrations in the text.
 \dsa\  of course reduces to flat space for $h=t=0$.  Another useful way to write flat space is in
 ``ingoing Rindler'' form\foot{Writing $\v =2\ln t^+,~~r=-t^+t^-$, the 2D part of the metric becomes $-4dt^+dt^-$.}
 \eqn\bcv{\eqalign{h(r)&=r,~~~~t=0,\cr ds_R^2&=-rd\v ^2+2d\v dr+dx^idx_i.}}
 Observers at fixed $r>0$ and $x^i$ are then Rindler observers.

 The asymptotically AdS$_{p+2}$ black p-brane
 solutions are
\eqn\ggl{\eqalign{h&={r^2 \over R^2}\Big(1-{r_h^{p+1}\over r^{p+1}}\Big),~~~e^t={r\over r_h},\cr
ds_{BB}^2 &=- {r^2 \over R^2}\Big(1-{r_h^{p+1}\over r^{p+1}}\Big)d\v ^2+2d\v dr+ {r^2\over r_h^2}dx_idx^i .}}

Rindler space \bcv\ is a limit of the black brane geometry \ggl. To see this define
\eqn\dap{\eqalign{r'&= {R^2(r-r_h) \over (p+1)r_h},\cr \v '&={(p+1)r_h \over R^2}\v .}}
The horizon is then at $r'\to 0$ near which
\eqn\saw{ds_{BB}^2=\left( -r'd\v '^2+2d\v 'dr'+dx^idx_i \right) \left(1+{\cal O}({r'\over R^2})\right).}
Hence Rindler space is both the near-horizon and $R\to \infty$ limits of the black brane.

If we add a $U(1)$ gauge field and charge density $Q$ to the black brane, the metric and gauge field $A$ are
\eqn\mvj{\eqalign{h&={r^2 \over R^2}\left(1-\left(1+\alpha Q^2\right){r_h^{p+1}\over r^{p+1}}+\alpha Q^2{r_h^{2p}\over r^{2p}}\right),~~~e^t={r\over r_h},\cr
A &= {Q r_h \over p-1}\left(1-{r_h^{p-1}\over r^{p-1}}\right)d\v.}}
Here $\alpha={R^2 8\pi G \over p (p-1)},$ and we have set the electromagnetic coupling constant to one.

A specific example with no AdS region is provided by the well-studied asymptotically flat NS5
brane (we omit the case of general $p$ for brevity). This is a solution of ten-dimensional supergravity with a three-form field strength threading an $S^3$  which surrounds the brane. This ten-dimensional geometry of course is not
of the form \dsa. However if we Kaluza-Klein reduce to 7 dimensions on the $S^3$, then it does take this form.  In the 7-dimensional Einstein frame and coordinates \dsa, the metric is
\eqn\dsefs{ h(r) = y^{6/5}\left(1+{L^2\over y^2}\right)^{1/5} \left(1-{y_h^2\over y^2}\right),\;\;
e^{2t(r)} = {y^{6/5}\over y_h^{6/5}} \left(1+{L^2\over y^2}\right)^{1/5}\left(1+{L^2\over y_h^2}\right)^{-1/5} }
Here $y(r) $ is the solution of
\eqn\dfe{r=\int^y dy'\; y'^{6/5}\left(1+{L^2\over y'^2}\right)^{7/10} .}
The right hand side is a hypergeometric function.

We wish to stress that our approach applies to geometries of the general form \dsa\ and is not tied to the above specific examples. Other interesting examples include the  proposed holographic duals to superconductors \HartnollKX, for which the metric cannot in general be found analytically, or other cases (\refs{\KlebanovHB,  \gmss} to mention a few) which are not asymptotically AdS and correspond to systems which are not conformally invariant in the UV.

\newsec{Electromagnetic warmup}
    In this section we warm up to the gravity problem by considering the conceptually similar,
    but mathematically simpler, problem of an electromagnetic field $F$ propagating
    in the geometry \dsa.

    \subsec{The setup}

    Our first step is to introduce a cutoff surface  $\Sigma_c$ outside the horizon in the general geometry \dsa
    \eqn\dfl{\Sigma_c:~~~~~~r=r_c>r_h,} with coordinates \eqn\cfg{x^a \sim (x^i,\v ).} The induced metric on $\Sigma_c$ is flat and given by \ykj\ with $r=r_c$.  We wish to study the dynamics of an electromagnetic field $F$ within the region
    \eqn\ggo{r_h\le r \le r_c.}
    This requires boundary conditions at both $r_c$ and $r_h$.  Since $r_h$ is a black hole horizon, we impose ingoing boundary conditions there.   At $r_c$, a natural Dirichlet-like choice is to fix the components of the field strength tangent to $\Sigma_c$
     \eqn\frav{F_{ab}(x^e,r_c)=f_{ab}(x^e)~~~~a,b,e=0,...p.}

We view the $f$s as the parameters defining the cutoff theory, and
$\Sigma_c$ as the place where experiments are set up
   and measurements made which probe the entire region \ggo\  below the cutoff.
   Fixing a radius where experiments are performed is dual, in the fluid picture, to fixing the scale at which experiments are performed.

  The boundary conditions at $r_c$ and $r_h$ do not uniquely specify a solution of the Maxwell equations for $F$. The problem is to describe the remaining dynamical degrees of freedom. We will see that, in the limits we consider, they are described by a single function $q(x^a)$ which obeys a simple diffusion equation on p+1-dimensional Minkowski space.  $q$ can be  thought of as a charge density on the horizon or, by a simple rescaling, as a charge density at the cutoff $q_c$. We will compute the diffusivity and also see that the data $f_{ab}$ specifying the boundary conditions at $\Sigma_c$ function as a source for the charge density. This enables cutoff observers to probe the dynamics of the charge density.

   Let us now turn to the details of this description.
\subsec{Equations and boundary conditions}
  We have a bulk gauge field with components $F_{\v r},~F_{ir},~F_{i\v },~F_{ij}.$
 In terms of these the bulk Maxwell's equations may be written
 \eqn\jkl{r:~~~~e^{2t}\p_\v F_{\v r}+\p^i F_{\v i}=h\p^iF_{ir},}
 \eqn\aio{i:~~~~\p_\v F_{ir}+((2-p)t' -\phi')F_{\v i}-\p_rF_{\v i}+h'F_{ir}=-h\p_rF_{ir}+h((2-p)t'-\phi')F_{ir}+e^{-2t}\p^jF_{ji},}
 \eqn\ssi{\v :~~~~\p_rF_{\v r}+(pt'+\phi')F_{\v r}+e^{-2t}\p^iF_{ir}=0,}
 where $\p^i=\delta^{ij}\p_j$ and for future utility we have allowed for a position-dependent gauge coupling ${1 \over g^2}=e^\phi$ normalized so that $g(r_h)=1$ and $\phi(r_h)=0$.
 In addition we will need the Bianchi identities
 \eqn\wwd{\p_rF_{i\v }= \p_\v F_{ir}-\p_iF_{\v r},}
 \eqn\bcl{\p_rF_{ij}=\p_iF_{rj}-\p_jF_{ri}.}

 We wish to impose ingoing boundary conditions on the gauge field at the future horizon.
 As our coordinates are regular on this horizon, we require $F$ to be regular there:\foot{We thank Stephen Green and Robert Wald for pointing out an error in the previous version of our paper.}
 \eqn\jju{F_{ri}(r_h)= {\rm finite}.}
 The other data at the horizon are the horizon current and charge density, defined as\foot{We may also write this in terms of the normal \dxw\ as $(q, j_i)^a = F^{ab}\ell_b$.}
 \eqn\jip{j_i(x^a)\equiv F_{i\v }(x^a,r_h),~~~~q(x^a)\equiv  F_{r\v }(x^a,r_h),}
 together with $F_{ij}(x^a,r_h)$.
 Given the regularity condition, the Maxwell equation \jkl\ at the horizon becomes current
 conservation
 \eqn\dst{\p_\v q+\p^ij_i=0.  }

\subsec{Long wavelength expansion}

In this subsection we introduce a non-relativistic long-wavelength expansion which is
suitable for studying hydrodynamics.

\noindent{$\bullet$\it ~~ Solving the equations}

The general solution of the Maxwell equation cannot be found analytically in a general geometry of the form
\dsa.
To proceed further we consider a long-wavelength expansion parameterized by $\epsilon\rightarrow 0$. We take temporal and spatial derivatives to have the non-relativistic scaling
\eqn\fpl{\p_\v \sim \epsilon^2,~~~~~\p_i \sim \epsilon. } The gauge field has the associated expansion
\eqn\wwn{\eqalign{F_{ir}&= \epsilon \bigl( F^0_{ir}+ \epsilon F^1_{ir}+\cdots \bigr), \cr
F_{\v r}&= \epsilon^2 \bigl( F^0_{\v r}+ \epsilon F^1_{\v r}+\cdots \bigr), \cr F_{ij }&= \epsilon^2\bigl( F^0_{ij}+ \epsilon F^1_{ij }+\cdots \bigr),\cr F_{i\v }&= \epsilon^3\bigl( F^0_{i\v }+ \epsilon F^1_{i\v }+\cdots \bigr). \cr}}
We will solve for $F^0_{\mu\nu}(x^a,r)$ in terms of its value at the horizon $r=r_h$ by integrating the first order radial evolution equations  \aio\ and \ssi\ outward to $r>r_h$, and then demanding agreement with the boundary conditions \frav\ when the cutoff $r=r_c$ is reached.
 At lowest order in $\epsilon$ these equations  are
\eqn\aizo{i0:~~~~h'F^0_{ir}=-h\p_rF^0_{ir}+h((2-p)t'-\phi')F^0_{ir},}
 \eqn\szsi{\v 0:~~~~\p_rF^0_{\v r}+(pt'+\phi')F^0_{\v r}+e^{-2t}\p^i F^0_{ir}=0.}
 There are no non-trivial solutions of \aizo\ which obey the ingoing boundary condition \jju.
  Therefore  \eqn\dse{F^0_{ir}=0.}The general solution of the second equation \szsi\ is then
 \eqn\rds{F^0_{\v r}(x^a,r)=-e^{-pt(r)-\phi(r)}q^0(x^a).}  The leading order Bianchi identity
 \eqn\ssa{\p_rF^0_{i\v }=e^{-pt(r)-\phi(r)} \p_i q^0}
 then implies the leading term in $F_{i\v }$
\eqn\ook{F^0_{i\v }(x^a, r  )=- \int_{r}^{r_c} ds e^{-pt(s)-\phi(s)} \p_i q^0(x^a) +f^0_{i\v }(x^a),}
where we have used the boundary conditions at the cutoff to determine the integration constants.
 Evaluated at the horizon $r=r_h$ \ook\  gives the Fick-Ohm law in $p+1$ dimensions
 \eqn\wwws{j^0_i=-D^{EM}_c\p_iq^0+f^0_{i\v },}
 with diffusivity given by the line integral
 \eqn\dia{D^{EM}_c(r_c)=\int_{r_h}^{r_c}ds e^{-pt(s)-\phi(s)}.}
 Current conservation \dst\ then implies
 \eqn\ssz{\p_\v q^0=D^{EM}_c\p^2q^0-\p^if^0_{i\v }.}
 In particular if we choose conducting boundary conditions so that the electric field vanishes at the cutoff we find Fick's second law
 \eqn\sszx{\p_\v q^0=D^{EM}_c\p^2q^0.}
 Taking the Fourier transform of this equation, we see that the charge density propagates according to the dispersion relation
 \eqn\dsp{i\omega=D^{EM}_c k^2.}

 We still need to solve for $F^0_{ij}(x^a, r)$.  The leading term in the Bianchi identity \bcl\ is
 \eqn\wao{\p_r F^0_{ij}=0.}
 The solution of this with the given boundary conditions is simply
 \eqn\poi{F^0_{ij}(x^a,r)=f^0_{ij}(x^a).}
 Hence there are no dynamics associated with $F^0_{ij}$.

   To leading order in $\epsilon$, \dse, \rds, \ook\ and \poi\ comprise the most general solution of the
 Maxwell equation with the given boundary data. The solution is characterized by a single function in $p+1$ dimensions, the charge density $q^0(x^a)$, which obeys the dispersion relation \ssz. Away from the scaling limit $\epsilon=0$, the dispersion relation \dsp\ for $q$ has corrections of order $k^4$.

 Other parameterizations of the solution space are possible. One alternative is to take the normal component of the electric field at the cutoff, which is not fixed by the boundary condition there. This is simply related at leading order to the horizon charge density, according to  \rds, by\foot{We could of course have couched our entire argument in terms of $q_c$, but the derivation is then burdened by extra terms which cancel in the end.  }
\eqn\dsx{q_c^0(x^a)\equiv e^{2 t(r_c)} F^0_{r\v }(x^a, r_c)=e^{-(p-2)t(r_c)-\phi(r_c)}q^0(x^a),}
and hence obeys the same dispersion relation.
This parameterization is more natural from our perspective as we want to think of measuring the dynamical modes and dispersion constants with experimental devices positioned on the cutoff surface $\Sigma_c$.
The charge density can be sourced, according to \ssz, by turning on an electric field $f_{i\v }$. Hence the dispersion constant can be measured by turning on a tangential electric field at the cutoff and watching how the normal component decays.  We shall henceforth
adopt $q_c$ as our basic variable.

\noindent{$\bullet$\it ~~ Normalizing the diffusivity }

  As it stands, the value of  $D^{EM}_c$ \dsp\ is not very meaningful because
 it is dimensionful and can be set to any value by a change of coordinates. We therefore introduce the proper frequency $\omega_c$ and proper momentum $k_c$ conjugate to proper time and distance
 in the cutoff hypersurface $r=r_c$. These are
 \eqn\xaz{\omega_c={\omega \over \sqrt{h(r_c)}}\sim -i{\p \over \p x_c^0},~~~k^i_c=e^{-t(r_c)}k^i\sim i{\p \over \p x_c^i}.}

 Observers at the cutoff equipped with a thermometer measure a non-zero temperature $T_c$.\foot{The formulation here naturally leads to a cutoff-dependent temperature. If we want to keep the temperature at the cutoff fixed
as we change $r_c$, we would have to simultaneously move in the space of black-brane solutions.} To determine $T_c$, note that very near the horizon $r\to r_h$,  observers with worldlines of fixed $r$ and $x^i$ are highly accelerated.  They therefore detect a Rindler temperature which
diverges as\foot{We have set $\hbar=1$, otherwise it would appear on the right hand side of this equation.}
 \eqn\tri{T_R={\p_rh\over 4\pi  h^{1/2}}={\sqrt{h'(r_h)}\over  4\pi \sqrt{r-r_h} } + {\cal O} \left(\sqrt{r-r_h}\right)}
 for any smooth quantum state. For an equilibrium state, the temperature as a function of $r$ is determined by the Tolman relation
 \eqn\dpu{T(r)\sqrt{h}=T_H={\rm constant}.}
 For an asymptotically flat spacetime, $h \to 1$ at infinity, and $T_H$ is the Hawking temperature.
The relation  \dpu\ together with the boundary condition \tri\ determines the local temperature at the cutoff to be
 \eqn\wes{T_c \equiv T(r_c)= {T_H \over\sqrt{h(r_c)}}, ~~~~~T_H= {h'(r_h) \over 4\pi }.}
 The quantity $\bar D^{EM}_c$ defined by
 \eqn\dsdp{i\omega_c={\bar D^{EM}_c \over T_c} k_c^2, }
then gives the diffusivity in units of the cutoff temperature.
Our final expression for the coordinate-invariant, dimensionless diffusivity $\bar D^{EM}_c$ is
 \eqn\dxzl{\bar D^{EM}_c={ e^{2t(r_c)} h'(r_h) \over 4\pi h(r_c)}\int_{r_h}^{r_c}ds e^{-pt(s)-\phi(s)}.}

For $r_c\to r_h$, $\bar D^{EM}_c$ has the universal behavior
\eqn\rpa{\bar D^{EM}_c={1 \over 4\pi }+{\cal O}({r_c-r_h} )}
for any geometry.  Unsurprisingly, it follows from \bcv\ that the leading term is exact for all $r_c$ for Rindler space.
For an $AdS$ black p-brane with $p > 1$ and a constant gauge coupling
($\phi=0$) the integral is easily evaluated and yields
\eqn\bds{\bar D^{EM}_c={p+1\over 4\pi (p-1)}{1-({r_h \over r_c})^{p-1} \over   1-({r_h \over r_c})^{p+1} }.}
Near the boundary $r_c=\infty$ we obtain
\eqn\rdst{\bar D^{EM}_\infty={p+1\over 4\pi (p-1)}+{\cal O}\left({r_h \over r_c} \right).}
The leading term agrees with the result of Kovtun and Ritz \KovtunKX\ and Starinets \SonVK.
We see that the dimensionless diffusivity runs from their result  to the universal $1 \over 4 \pi $ in \rpa\ as the cutoff is taken from infinity to the horizon.

\subsec{Near horizon expansion}
Here we consider the near-horizon dynamics when $r\rightarrow r_h$. There is then no need to make the long-wavelength $\epsilon$ expansion which required small $\omega$ and $k$. Inspired by the diffusion behavior of \dsp, or equivalently \dsdp, we let
$\left( \tau, r \right) \rightarrow \left(t, \rho\right)$ where
\eqn\taueqn{\tau = {t \over \lambda},~~~~r = r_h+\lambda \rho.}
Taking the limit $\lambda \rightarrow 0$ corresponds to focusing on the near-horizon region. We also have
\eqn\hreqn{h(r) = \lambda \rho + {\cal O}(\lambda^2), ~~~~ e^{t(r)} = 1 + {\cal O}(\lambda).}
Since we wish to consider couplings which are regular in $r$ we can write
\eqn\ephieqn{e^\phi= e^{\phi_0}+\cal{O}(\lambda)}
where $\phi_0$ is a constant. The gauge field has an associated expansion
\eqn\fmunuexpansion{F_{\mu\nu}=F_{\mu\nu}^{(0)} + F_{\mu\nu}^{(1)} \lambda + {\cal O}(\lambda^2).}
At leading order \aio\ and \ssi\ give
\eqn\firhozeroth{F_{i\rho}^{(0)}=0, ~~~~\p_\rho F_{t\rho}^{(0)}= 0.}
From \aio\ and \wwd\ we find
\eqn\fti{\p^iF_{i\rho}^{(1)}= \p^2 F_{t\rho}^{(0)}, ~~~~~F_{ti}^{(0)}= \left(\rho-\rho_c\right) \p_i F_{t\rho}^{(0)} + f_{ti}(t,x^j)}
where $f_{ti}$ is any function of $(t,x^i)$. The cutoff surface is at $r_c=r_h + \lambda \rho_c$ and there we impose the boundary condition $F_{i t}^{(0)}=0$. Identifying $F_{t\rho}^{(0)}=e^{-\phi_0} q_c$ together with equation \jkl\ at $\cal{O}(\lambda)$ leads to
\eqn\ftrhozeroth{\p_t F_{t\rho}^{(0)} = \rho_c \p^2 F_{t\rho}^{(0)}}
which is equivalent to
\eqn\qceqn{\p_t q_c = \rho_c \p^2 q_c.}
Alternatively we can write this in terms of the dispersion relation
\eqn\rypp{i \omega_c = {1 \over 4 \pi T_c} k_c^2}
with {\it no subleading $k^4$ corrections for $r_c \rightarrow r_h$}. This makes sense since taking the cutoff to the horizon should correspond to an infrared limit in which higher derivative corrections are scaled away.

\subsec{Subsummary}
   To summarize this section, for large $r_c$, which should correspond to taking the fluid cutoff to the UV,  the charge density obeys a complicated non-linear dispersion relation. For small $\omega \sim k^2$, Fick's law holds (for any $r_c$) with a dimensionless diffusivity $\bar D_c^{EM}$ expressed as a radial line integral from the horizon to the cutoff.  In the infrared limit as the cutoff is taken to the horizon, the dispersion relation simplifies to the Fick law and $\bar D_c^{EM}$ takes the universal value ${1 \over 4 \pi}$.

 \newsec{Gravity }

 Now we will adapt this to gravity.  The linearized Einstein equations become rather complicated in the general background \dsa\ but have been completely analyzed in a series of papers whose results we shall use without rederivation.
 One important feature, demonstrated in \ks\ and \KodamaFA, is that the linearized gravity fluctuations decouple into so-called vector, scalar (or sound) and tensor type, characterized by their transformations under the $O(p)$ rotational symmetry. We consider these different types in turn.

 We stress that throughout this section we assume that linearized metric fluctuations
 are governed by equations of motion whose form is given by the linearization of the Einstein tensor. $R^2$ corrections or even certain matter couplings could
 change this. Quite often these equations are unchanged, especially for the shear mode.  We expect qualitative generalizations of our results to hold for any type of couplings.

 \subsec{Vector/shear modes}

In this section we consider the vector fluctuations, which turn out to be the most interesting for our purposes. These have nonzero $h_{ia}$ components which obey
\eqn\ccx{\p^ih_{i \v }=0,\quad h_{ij}=F(r)\p_{(i} h_{j)\tau}.}
In terms of these we define U(1) gauge fields
 \eqn\ppl{ A^{i  }= e^{-2t} h^i_a dx^a,} and field strengths
 \eqn\kkj{F^i=dA^i. }
 Here the index $i$ labels the different field strengths.
Following \KovtunWP, by considering dimensional reduction along the polarization direction of the metric perturbations,  the linearized Einstein equation for each $F^i$ is precisely that of an abelian gauge field with  position-dependent gauge coupling
 \eqn\ktr{e^\phi=e^{2t}.} This reduces the equation for the metric vector fluctuations to  $p-1$ independent copies of the Maxwell equations. We may therefore read the answer off of the solution of the previous section.  We take ingoing boundary conditions at the horizon and Dirichlet boundary conditions on $h^i_a (x^a,r_c)$ at the cutoff.
 This amounts to fixing the induced metric on $\Sigma_c$.
The analog of the charge density $q_c$ is the vector field
 \eqn\oos{v^i(x^a)\equiv (\p_r -2 t')h^i_{\v} (x^a,r_c).}
 It follows from \ccx\ that these are divergence free
 \eqn\iws{\p_i v^i=0.}
 The analog of the Fick-Ohm law is
  the forced linearized Navier-Stokes equation
   \eqn\ddx{\p_\v v^i=D_c \p^2 v^i +s^i}
where the diffusivity is
 \eqn\saz{D_c=\int_{r_h}^{r_c}ds e^{-(p+2)t(s)}.}
 Here the forcing term $s_i$ is
 \eqn\wws{s^i=e^{-2t(r_c)}(\p_j\p_\v h^i_j(x^a,r_c)-\p^2h^i_\v (x^a,r_c)).}

We wish to define a dimensionless coordinate invariant diffusivity\foot{The quantity $\bar D_c$,  which is the kinematic viscosity times the temperature, differs from the dynamic viscosity $\eta$ by a factor involving  the energy density, temperature and pressure. It is a more basic quantity from our perspective in that it is directly related to the measured decay rate of a perturbation. At this point we cannot define $\eta$, because we have not determined the energy density or pressure. This requires a bit more machinery - the Brown-York stress tensor - and will be worked out in section 5.} $\bar D_c$ by transforming to proper coordinates and multiplying by the local temperature $T_c$.  This proceeds exactly as in the electromagnetic case and the result is
 \eqn\zl{\bar D_c={ e^{2t(r_c)} h'(r_h) \over 4\pi h(r_c)}\int_{r_h}^{r_c}ds e^{-(p+2)t(s)}.}
 When the boundary conditions are chosen so that the forcing term vanishes,
 the transverse velocity fields $v^i$ propagate with the dispersion relation\foot{We have set $\hbar =c =1$, and the Newton constant $G$ does not enter our calculations. In more general conventions $i\omega_c=\hbar c^2 \bar D_c k_c^2$. The $\hbar$ appears in this classical calculation because the diffusivity is expressed in terms of the local temperature which is itself proportional to $\hbar$. }
 \eqn\dsp{i\omega_c=\bar D_c k_c^2.}

 The dimensionless diffusivity behaves universally in the infrared as the cutoff is taken to
 the horizon\foot{The results of \refs{\KatsMQ,\BriganteNU , \BuchelVZ \cremtwo,\MyersIJ,\cremone} suggest the universal value will be modified by higher-derivative gravity corrections.}
 \eqn\sei{\bar D_c \to {1 \over 4 \pi} ~{\rm for}~ r_c \to r_h.}
 This agrees with the result obtained three decades ago in \thesDamour. It applies for any geometry of the form \dsa. It is really a property of the linearized Einstein equation in Rindler space, which is the only relevant region for the calculation in the $r_c\to r_h$ limit.

 \subsec{A D-theorem and other special properties of $\bar D_c$}

 So far the discussion of the electromagnetic and gravitational diffusivity have been exactly parallel. However some special features arise in the gravitational case. This first is that the integrand of \saz\ turns out to obey
 \eqn\vct{e^{-(p+2)t}=\partial_r\left[{h e^{-(p+2)t}\over h'-2t'h}\right]+16\pi G\left[ {h e^{-(p+2)t}\over\left(h'-2t'h\right)^2}\right] T_{\mu\nu}\zeta^\mu\zeta^\nu.}
 Here $T_{\mu\nu}$ is the bulk matter stress tensor and $\zeta$ is any null vector tangent to the brane with time component $h^{-1/2}\p_\tau$.
 Hence if there is no matter or if $T_{\mu\nu}\zeta^\mu \zeta^\nu=0$, the integrand is a total derivative.  $\bar D_c$ is then given by the simple expression
 \eqn\roiu{\bar D_c={h'(r_h)\over 4 \pi}{e^{-pt(r_c)}\over
h'(r_c)-2t'(r_c)h(r_c)}.}

The fact that the expression for the diffusivity can be integrated stems from the fact that the shear modes are pure gauge at zero momentum (we generalize here \refs{\SonVK}).  $h_{i\tau}$ is nonzero for $k^i=\omega=0$, and we can derive its radial dependence from the zero momentum Einstein equation.
However we can also solve this equation with a gauge transformation of the form
\eqn\gtrf{\delta x^i= e^i \tau +B e^i \int e^{-2t} dr ,~~~~~ \delta \tau =- B e_ix^i }
 with $B,~e^i$ arbitrary  constants. The second term in $\delta x^i$ is added to preserve our gauge condition $h_{r\mu}=0$.   The nonzero component is
 \eqn\plk{h_{i\tau}= \left(Bh + e^{2t}\right) e_i }
 In order to preserve the boundary condition $h_{i\tau}(r_c)=0$, we must take
 \eqn\ruk{B=-{e^{2t(r_c)} \over h(r_c)},}
 resulting in  \eqn\zmp{h_{i\tau}(r) =  e_i \bigl(e^{2t(r)}-{h(r)\over h(r_c)}e^{2t(r_c)}\bigr)\equiv f_1(r)e_i. }
 Of course we already found this solution, which is not pure gauge at non-zero-momentum,  using the Einstein equation. The present derivation has the distinct advantage that it follows from gauge invariance and therefore is completely independent of the prescribed dynamics.  Note that we can also find matter perturbations compatible with this gravity solution by the same method.

 $\bar D_c$ can be expressed as a  ratio of the coefficients of $\p_r h_{i\tau}$ and $\p_r h_{ij}$. The latter vanishes at zero momentum so we need to work to next order in $\epsilon$ expansion. We promote $e^i\rightarrow e^i(x^a)$, with $\p_i e^i(x^a)=0$. By symmetry $h_{ij}= e^{2t}f_2(r)(\p_ie_j+\p_je_i)$ for some function $f_2(r)$, determined from the $ij$ components of the linearized Einstein equation
 \eqn\ije{ \delta G_{ij} = 8\pi G \delta T_{ij}.}
Again by symmetry the variation of $T _{ij}$ with the matter fields is of the form $t_m(\p_ie_j+\p_je_i)$ and \ije\ reduces to a single equation which becomes
\eqn\kije{e^{-(p-2)t}\p_r{[-e^{(p-2)t}f_1+e^{tp}h\p_r f_2  ]}  = -16\pi G t^m}
in the long wavelength limit.
In many cases (all that we have studied) $t_m=0$ and hence
\eqn\mkije{h(r)\p_rf_2(r) = e^{-2t(r)}f_1(r) - f_1(r_h) e^{-pt(r)}.}
This is then enough information to reproduce \roiu.  So the universality of \roiu\ is at least in part a consequence of zero-momentum gauge invariance.

  Interestingly, if we plug in the neutral AdS black brane metric \ggl\ we find the surprising result
 \eqn\rik{\bar D_c= {1 \over 4 \pi} }
 for any value of $r_c$. Hence in this particular case $\bar D_c$ (unlike $\bar D^{EM}_c$ for the same geometry) does not run. In general one expects
 $\bar D_c$ and as well as all transport coefficients to run. Indeed for the charged black brane \mvj\ we will exhibit a rather nontrivial RG flow in section 4.5 below. The constant value appearing in \rik\ is directly related to the fact that ${\eta \over s}$ does not run, which will be discussed in section 6 below. However for now we briefly note the relation.  The viscosity is related to $\bar D_c$ by the relation $\eta={\bar D_c ( {\cal E}+{\cal P})\over T_c }$ where ${\cal E}$ (${\cal P}$) is the energy density (pressure). In the absence of chemical potentials, the entropy density is determined from
 $T_c s = {\cal E}+{\cal P}$, implying $\bar D_c={\eta \over s}$. When there are chemical potentials such as for the charged black brane this simple formula no longer holds, and $\bar D_c$ runs.

We now show that, in a wide range of circumstances,  $\bar D_c$ decreases with increasing $r_c$, i.e. it  increases under RG flow into the infrared. We follow in spirit the A-theorem of \FreedmanGP. Define the quantity
\eqn \form{ H(r_c)\equiv e^{(p-2)t(r_c)}h^2(r_c)\p_{r_c} \Big({4\pi\bar{D}_c \over h'(r_h)}\Big)=e^{-2t}h-(h'-2ht')e^{pt}\int_{r_h}^{r_c}ds e^{-(p+2)t(s)}.}
Since $h(r_h)=0$, we have $H(r_h)=0$.  The gradient of $H$ then obeys
\eqn \formm{\p_r H(r) = -16\pi G T_{\mu\nu}\zeta^\mu\zeta^\nu e^{pt(r)}\int_{r_h}^{r}ds e^{-(p+2)t(s)}.}
This relation employs the Einstein equation
\eqn\posit{ 16\pi G T_{\mu\nu}\zeta^\mu\zeta^\nu= 2G_{\mu\nu}\zeta^\mu\zeta^\nu=h''+(p-2)h't'-2 t''h-2pt'^2h. } The null energy condition implies $T_{\mu\nu}\zeta^\mu\zeta^\nu \ge 0$, and hence that
$\p_r H \le 0$.    Since $H(r_h)=0$ we conclude that $H(r)\leq 0$ for $r\geq r_h$. It then immediately follows from the definition \form\ that
\eqn\grav{\p_{r_c} \bar{D}_c\leq 0.}
We expect that the general approach of this proof will be applicable to a wide range of situations. However, we wish to note that the precise result \grav\ may also be obtained from \roiu\ simply by differentiating with respect to $r_c$.

 \subsec{Tensor modes}

  The tensor modes of the metric have nonzero components $h_{ij}$ with $h_i^{~i}=\p_i h^i_j=0$.  The equations governing their fluctuations appear in \ks, and are equivalent to a $(p+2)$-dimensional scalar Laplacian for each of the $(p-2)(p+1)/2$ tensor components.  The analysis of these modes  in our setup  is very similar to the one given in section 3 for the $F_{ij}$ modes of the electromagnetic fluctuations, and will not be spelled out here. The conclusion is that, in either the near-horizon or long-wavelength expansions \fpl, their values are everywhere fixed by the boundary conditions on $\Sigma_c$ and the horizon. There are no dynamical modes in these expansions.

 \subsec{Scalar/sound modes}
 In this subsection we consider the scalar or sound mode.  The equations of motion are somewhat complicated and we restrict ourselves here to the case of the AdS black threebrane so that $p=3$, although we expect the more general case to be similar.  For $p=3$ the equations have been fully analyzed in \ks. They found that the metric fluctuations are determined by a certain linear combination of the components, denoted $Z_2$, obeying a second order radial differential equation which they express (equation 4.35) in Schwarzschild-like coordinates.
 For our purposes it is more convenient to work in the advanced coordinates \ggl, because  these are smooth at the future horizon and the ingoing boundary condition is simply regularity at $r=r_h$.  $Z_2$, which is a Fourier transform with respect to Schwarzschild time, is traded in these coordinates for the Fourier transform with respect to $\tau $ denoted here by $X$. $X$ is related to $Z_2$ via a factor of $e^{i \omega \int^r{dr' \over h}}$:
 \eqn\iaz{X=\left( {r-r_h \over r+r_h}\right)^{{-i\omega R^2 \over 4 r_h}}Z_2 }Rewritten in terms of $X$ and the coordinate $r$ the equation for the sound mode,
 equation (4.35) of reference \ks , is
 \eqn\gasc{\eqalign{& \left(r^4-r_h^4 \right)\p_r^2 X  + \left(r^4-r_h^4 \right) \left[ {5 r^4-r_h^4 \over r \bigl (r^4 -r_h^4 \bigr)}+{8 r_h^4 k^2 \over r r_h^4 k^2 +3 r^5 \bigl(-k^2+\omega ^2\bigr)}+{ i R^2  \omega  \over r^2-r_h^2}\right] \p_r X  =\cr & \left[k^2 R^4 -{\omega^2 R^4 \bigl(3 r^2 + r_h^2\bigr) \over 4 (r^2 + r_h^2)} -{i \omega R^2 \bigl(3 r^2+r_h^2\bigr)   \over 2 r}+ {4 r_h^4 k^2 \bigl(4 r_h^4-i \omega R^2  r \bigl(r_h^2+r^2\bigr)  \bigr) \over r^2 (r_h^4 k^2+3 r^4 (-k^2+ \omega^2))} \right] X.}}
 %

 Let us now analyze this equation in the long-wavelength expansion \fpl\ with $\omega\sim k^2 \sim
 \epsilon^2 \to 0$. Note that there are no poles appearing in \gasc\ for small $\omega \sim k^2$, so we can safely take $\epsilon \to 0$ for all $r$. One finds the equation for the leading term in the
 $\epsilon$-expansion of $X$
 \eqn\hadl{\p_r^2 X^0+ \left({5 r^4 - r_h^4  \over r (r^4 - r_h^4) } - {8 r_h^4 \over r (3 r^4 - r_h^4)}\right)\p_r X^0+ {16 r_h^8 \over r^2 (3 r^4 - r_h^4) (r^4 - r_h^4)} X^0=0,}
 which does not depend on $\omega$ or $k$. This has a unique solution, up to an overall scale,  which is
 non-singular at $r=r_h$. The scale is then fixed by the boundary condition at the cutoff $r_c$. Unsurprisingly we learn that there are no dynamical degrees of freedom in the sound mode in the expansion \fpl.

 It is interesting, but outside the scope of this paper, to consider an alternate scaling limit
 in which $\omega \sim k$. In this limit nontrivial sound modes may appear with some fixed velocity $v_s$. In the limit \fpl, all fixed velocities including $v_s$ are sent to infinity.

 Now let us now consider the near horizon expansion. In this expansion we take $r_c\to r_h$
 without any preassumed  relation between $\omega$ and $k$. In this  case it is not so simple to take $r_c\to r_h$,  as a quick  inspection of \gasc\ indicates there may be poles at
 \eqn\reat{\omega=\pm \sqrt{2 \over 3}k.}
 Let us first consider the case where $\omega$ does not take the value \reat.
 Then we can safely take $r_c\to r_h$, and as in our long-wavelength expansion above
 there are no degrees of freedom. Hence the only possibility for dynamical modes are those that obey \reat.  Equation \reat\ can be written in terms of the coordinate-invariant proper quantities and local temperature as
  \eqn\reat{\omega_c =\pm \sqrt{2\over 3}{ e^{t(r_c)} T_c \over T_H}  k_c.}
  This is a dispersion relation for a sound mode with velocity of sound
  \eqn\sav{v_s= \sqrt{2 \over 3}{e^{t(r_c)}T_c \over T_H}.}
 However note that as $r_c\to r_h$, $T_c$ and hence the velocity of sound goes to infinity.
 Hence, in an expansion in $r_c-r_h$, no sound modes appear and the fluid at the horizon is incompressible.

 This again is consistent with the expectation that the limit $r_c\to r_h$ is a nonrelativistic, low energy limit.  It is low energy because of the high redshifts, and non-relativistic because the degeneracy of the induced metric on $\Sigma_c$ appears only in the temporal and not the spatial components. In its most general form, the Navier-Stokes equation for a fluid contains both sound and shear modes. However one may take a further limit of these equations of the form \fpl\ in which velocities scale as $v^i\to \epsilon  v^i\to 0$ (see \BhattacharyyaKQ\ for a nice discussion). In this limit the sound velocity goes to infinity. The fluid retains its nonlinear interactions  and is described by the incompressible Navier-Stokes equation. The near-horizon limit resembles a bulk version of this limit.

 This leads us to the interesting conclusion that the fluid which lives at the horizon is, at linear order,  universally given by an incompressible fluid with dimensionless diffusivity $\bar D_c={1 \over 4 \pi}$.
\subsec{Charged black brane}

In this section we consider the case of the charged black brane geometry \mvj\ as a somewhat more non-trivial illustration of our approach.  If symmetry allows,
 a metric perturbation can source a matter perturbation already at the linear level.
For the charged black brane there is a matter perturbation of
the general form
\eqn\chaa{\delta A=a_j dx^j.} We can solve the coupled equations by transforming them back to the previous case.  First, let us define a shift of the ``gravitational'' field strength appearing in \ppl\ and \kkj\ by
\eqn\cha{\tilde{F}^j_{\tau r}=F^j_{\tau r}-16\pi G a^j e^{-2t}A_\tau '(r),}
with $A_\tau$ given in equation \mvj. (Note that, in a notational clash,  $neither$ $F$ nor $\tilde F$ here is the field strength of $A$! Instead, $F$ refers to the ``gravitational'' field strength as defined in \ppl\ and \kkj.) $\tilde{F}_{\tau r}^j$ then obeys exactly the same equations \jkl\
through \ssi\ obeyed by $F_{\tau r}^j$ in the neutral case.

Consequently, following the logic in Section 4.3, we find at lowest order, similarly to \rds,
\eqn\chb{\tilde{F}^j_{\tau r}=-e^{-(p+2)t}\tilde{v}^j(x^a)}
where
\eqn\chc{\tilde{v}^j(x^a) \equiv \tilde{F}^j_{r \tau}(x^a,r_h).}
Note that the effective ``charge'' $\tilde{v}^j$ in \chc\ is modified from the corresponding equation \jip.

Now we must examine the two remaining equations: the lowest order Bianchi identity
\eqn\bian{\p_r F_{i\tau}^j=-\p_i F_{\tau r}^j}
and the Maxwell equation for the gauge perturbation $a^j$
\eqn\gaugez{F_{\tau r}^j={1\over Q}\p_r\left(e^{(p-2)t}h \p_r a^j\right).}
Note that we have written these two equations in terms of $F_{\tau r}^j$; by doing so we can now quickly see that the right hand side of \bian\ is just a total derivative.  Thus we find, similarly to \fti\ for $r=r_h$,
\eqn\ookg{F_{i\tau}^j\left(x^a,r_h\right)={1\over Q} \int^{r_{c}}_{r_h} dr \; \p_r\left(e^{(p-2)t}h\p_i\p_r a^j\right)+f_{i\tau}(x^a).}

Before we can use \ookg\ to find the diffusivity $D_c$, we must first find $a^j$.  To do so, we simply plug \chb\ and \gaugez\ into \cha, additionally imposing the boundary condition $a^j(r_c)=0$.  We thus find
\eqn\chaz{a^j(r)={\tilde{v}^j(x^a) \over 16 \pi G Q}\left[1-e^{pt(r)-pt(r_c)}{h'(r)-2t'(r)h(r)\over h'(r_c)-2t'(r_c)h(r_c)}\right].}
Again we can find a dimensionless coordinate invariant diffusivity
which is
\eqn\che{\eqalign{\bar D^Q_c&={h'(r_h)\over 4 \pi}{e^{-pt(r_c)}\over
h'(r_c)-2t'(r_c)h(r_c)}\cr
&={p+1-\alpha Q^2 \left(p-1\right)\over 4\pi \left[p+1+\alpha Q^2
\left(p+1-2p{r_h^{p-1}\over r_c^{p-1}}\right)\right].}}}
Note that the first line of \che\ is identical to \roiu, corroborating the universality of \roiu.
However when we plug in the metric coefficients for the charged black brane to get the second line of \che, we see that, unlike the case of the neutral black brane, the diffusivity is no longer constant.
\subsec{Asymptotically flat black p-brane}
In our next example we consider the asymptotically flat $S^3$-compactified black NS5 solution \dsefs.   The 7-dimensional effective action for the rotationally invariant modes is Einstein gravity plus scalars, hence according to section 4.2, $\p_{r_c} \bar{D}_c=0$,  and $\bar{D}_c={\eta \over s}={1 \over 4 \pi}$  at any scale.  Of course this can be reproduced by direct calculation. The  local temperature is a non-trivial function of radial position:
\eqn\therm{ T_c={ 1\over 2\pi y_h y_c^{3/5}} \left(1+{L^2\over y_h^2}\right)^{-1/2} \left(1+{L^2\over y_c^2}\right)^{-1/10} \left(1-{y_h^2\over y_c^2}\right)^{-1/2}.}
General formulae from the following section for the energy density plus pressure give
\eqn\edpr{{\cal E}+{\cal P} = { y_h^2\over 8\pi G y_c^{18/5} } \left(1+{L^2\over y_c^2}\right)^{-3/5} \left(1-{y_h^2\over y_c^2}\right)^{-1/2}}
and entropy density
\eqn\entrd{ s = {e^{-5t(r_c)}\over 4G}= {y_h^{3}\over 4G y_c^{3}} \left(1+{L^2\over y_c^2}\right)^{-1/2}\left(1+{L^2\over y_h^2}\right)^{1/2}.}
We note that ${\cal E}+{\cal P}=T_c s.$
The $L\to \infty$ throat limit of these expressions should describe the thermodynamics of the quantum theory on the NS5-brane, but we will not further pursue this here.

 \newsec{Brown-York stress tensor}

 Up until this point, all of our results have followed from equations of motion and boundary conditions. These are sufficient to determine diffusion rates.  However they are not sufficient to determine quantities such as the energy, entropy or viscosity which depend on the normalization of the action.  In this section we introduce the Brown-York stress tensor $t_{ab}$
 whose overall coefficient depends on the action. We propose that, with our boundary condition,
$t_{ab}$ evaluated at the cutoff is the stress tensor of the fluid.  As we will see, the introduction of $t_{ab}$ makes the  calculations and analysis considerably more concise, as well as enabling us to compute more quantities.

The Brown-York stress tensor associated to a hypersurface $\Sigma$ with unit normal $n$ is defined  by
\eqn\ssz{t_{ab}={1 \over 8\pi G} \left(\gamma_{ab}K-K_{ab}+C\gamma_{ab} \right) ,}
where \eqn\jjuu{\gamma_{ab}\equiv g_{ab}-n_an_b}
is the induced metric on  $\Sigma$,  the extrinsic curvature is
\eqn\tth{\eqalign {K_{ab}& \equiv \half \cl_n \gamma_{ab}\cr &=\half\bigl( n^c\p_c\gamma_{ab}+\p_a n^c\gamma_{cb}+\p_bn^c\gamma_{ac}\bigr)\cr &=\half\bigl( \cl_ng_{ab}-n_an^c \cl_ng_{cb} -n_bn^c \cl_ng_{ac}\bigr) ,}}and $\cl_n$ is the Lie derivative along $n$.
$t_{ab}$ is ambiguous up to a constant multiple of the induced metric which we denote here $C$. This is equivalent to the ambiguity in shifting the pressure of a fluid by a constant.

 The Einstein equation $G_{\mu\nu}=8\pi GT_{\mu\nu}$ implies that  $t_{ab}$ obeys the conservation
law
\eqn\was{\nabla^at_{ab}=n^\mu T_{\mu b},}
where $a,b$ are projected to $\Sigma$, $\nabla$ is the  $\gamma$-covariant derivative and
$T_{\mu\nu}$ is bulk stress tensor.
We are interested in the case for which $\Sigma=\Sigma_c$ is the cutoff hypersurface
and will always impose ``energy-conserving'' boundary conditions so that the right hand side of \was\ vanishes. We then have \eqn\wxas{\nabla^at_{ab}|_{\Sigma_c}=0.}

Note that if we were to allow fluctuations in both the induced metric and extrinsic curvature, the linearization of \wxas\ about the background would be
 \eqn\wxas{\left[(\delta \nabla^a) t^0_{ab} +\nabla^{0a} (\delta t_{ab} )\right]_{\Sigma_c}=0.}
 This equation implies that the linearized Brown-York stress-tensor is $not$ in general conserved in the background geometry, precluding a simple interpretation as the linearized fluid stress tensor. However with our boundary conditions the induced metric on $\Sigma_c$ is held fixed and all is well.

\subsec{Background stress-energy}
Now we evaluate $t_{ab}$ for
our general metric \dsa\ at the cutoff hypersurface $\Sigma_c$. The unit normal is  \eqn\jjd{n={ \sqrt{h}}\p_r+{1 \over \sqrt h }\p_\tau} and
   \eqn\ssz{K_{\mu\nu}dx^\mu dx^\nu={ \sqrt h}\left[ -{h' \over 2}\left(d\tau-{dr \over h}\right)^2+t'e^{2t}dx_idx^i \right],}
   \eqn\ssdz{\gamma_{\mu\nu}dx^\mu dx^\nu=\left[ -h\left(d\tau-{dr \over h}\right)^2+e^{2t}dx_idx^i\right].}
   In intrinsic coordinates to $\Sigma_c$ as defined in \dio\

   \eqn\ssz{K_{ab}dx^a dx^b= -{h' \over 2{ \sqrt h}}(dx^0_c)^2+{ \sqrt h}t'dx_{ci}dx_c^i ,}
   \eqn\ssdz{\gamma_{ab}dx^a dx^b=\eta_{ab}dx_c^adx_c^b.}

  The leading order Brown-York stress tensor is
  \eqn\byg{t^0_{ab} dx^a dx^b={{ \sqrt h}\over 8\pi G}\bigl(-pt'(dx^0_c)^2+\bigl((p-1)t'+{h' \over 2 h }\bigr)dx_{ic} dx^i_c\bigr) + C'\eta_{ab}dx_c^a dx_c^b , }
  where all $r$-dependent quantities are evaluated at $r=r_c$. This is the stress tensor of a fluid
 at rest  with constant pressure $\cal P$ and  energy density $\cal E$.  The constant part of the difference $ {\cal E}-{\cal P}$ depends on the choice of constant $C$: the behavior of the linearized fluid depends only on the sum  \eqn\dsx{{\cal E}+{\cal P}={ {\sqrt h}\over 8\pi G}\left({h' \over 2 h }-t'\right).}
  Note that for $r_c\to r_h$, ${\cal E} +{\cal P}\to {T_c \over 4 G}=T_c s$, where $s$ is the entropy density of the horizon.

  \subsec{Perturbations}
  Now we consider a small perturbation $h$. It is convenient to choose the gauge
  \eqn\wwwa{h_{rr}=h_{r\v}=h_{ri}=0,}
  so that the metric retains the general form \dsa. Moreover, our boundary condition implies
  \eqn\fas{h_{ab}(r_c)=0.}  It then follows that the leading correction to the extrinsic curvature in the long wavelength limit simplifies to $2 K^1_{ab}=\sqrt{h}\p_r h_{ab} $, and
 \eqn\byzg{t^1_{ab}(x^a, r_c)={{ \sqrt h}\over 16\pi G}\bigl(-\p_rh_{ab} +\gamma_{ab}\gamma^{cd}\p_rh_{cd}\bigr) .}
 For the shear mode, only $h_{i\tau}$ and $h_{ij}$ are nonzero. Moreover, as shown in \KodamaFA\ the Einstein equation implies they are related by $F \p_{(j}h_{i)\tau}=h_{ij}$ for some function $F(r)$, as in equation \ccx\ previously. Conservation of the Brown-York stress tensor implies that $F(r_c)={\bar{D}_c \over T_H}$ on $\Sigma_c$ and
 \eqn\swp{ \p_r h_{ij}={\bar{D}_c \over T_H} \p_r \p_{(i} h_{j)\tau}.}
 Defining the velocity field
 \eqn\rto{v_i=-{e^{-t} \over 16 \pi G ({\cal E}+{\cal P})}\p_rh_{i\tau},}
 The linearized stress tensor can be written in intrinsic coordinates
 \eqn\byzga{t^1_{ab} dx_c^a dx_c^b=2({\cal E}+{\cal P})v_idx_c^id\tau_c  +\eta \p_iv_j dx_c^idx_c^j  }
with viscosity given by
 \eqn\fdr{\eta={\bar D_c  ({\cal E}+{\cal P})\over T_c}.}
 Using \wes, \roiu\ and \dsx,  this becomes
 \eqn\wwsax{\eta(r_c)={e^{-pt(r_c)} \over 16 \pi G},}
 in agreement with \thesDamour\ for $r_c \to r_h$ where $t\to 0$.

We see from the above that the stress tensor for the  linearized shear mode indeed takes the form of an incompressible fluid.
\newsec{RG flow, the first law of thermodynamics and $\eta\over s$ universality}

In this section we want to show that the first law of thermodynamics together with isentropy
of the radial flows is  equivalent to a radial component of the Einstein equation, and moreover imply that ${\eta \over s}$ does not run.  A general type of  equivalence between the Einstein equation and the first law has been demonstrated by Jacobson \theo, see also \PadmanabhanVY.
We suspect our equivalence is related to this - as well as the recent work \verl\ - but we defer this issue to future consideration.

For the present purposes, it is convenient to consider a quotient of the general geometry \dsa\ under shifts of $x^i$
\eqn\fik{x^i\sim x^i+n^i,~~~~n^i \in Z.}
This turns the spatial $R^p$ in the $p$-torus $T^p$ with $r$-dependent volume
\eqn\fsi{V_p=e^{pt}.} The total entropy $S=sV_p$ as a function of the total energy $E={\cal E}V_p$, pressure $\cal P$, charge $Q$, chemical potential $\mu$, inverse temperature $\beta=T_c^{-1}$ and volume $V_p$ are related by
\eqn\ffx{S=\beta E+\beta {\cal P}V_p-\beta \mu Q.}
In general there will be one $\mu Q$ term for each chemical potential present; in our case we consider only the charged brane given in \mvj, for which $\mu=A_\tau/\sqrt{h}$.

Now let us consider the first law of thermodynamics for radial variations under the assumption that the variation is isentropic.  This reads
\eqn\read{0=\p_rS=\p_r\beta \left( E+{\cal P}V_p- \mu Q \right)+\beta \left ( \p_rE+ \p_r{\cal P}V_p+ {\cal P}\p_rV_p-\p_r\mu Q\right).}
Using expressions \wes, \dsx\ and \fsi\ for the thermodynamic quantities\foot{We are considering here only the thermodynamic  quantities associated to the background. It would be interesting, but beyond the scope of this paper, to understand if this reasoning applies at quadratic order in the fluctuations, where shear dissipation produces entropy.} one finds \read\ becomes
\eqn\rew{\p_r S = {V_p \over 16\pi G T_H}(h''+(p-2)h't'-2t''h-2pt'^2h)-{V_p\over T_H} T_{\mu\nu}\zeta^\mu\zeta^\nu.}
Here we have also used $V_p T_{\mu\nu}\zeta^\mu\zeta^\nu=Q^2 e^{-pt}$ for the charged brane. Comparing with \posit, we see that the right hand side is exactly a component of the Einstein equations. Therefore, isentropy of the RG flow implies a radial Einstein equation. Of course this can be turned around to state that the radial Einstein equation implies the isentropy of RG flow.

Now if $S=constant$, then
\eqn\tru{s={e^{-pt } \over 4 G},}
where the overall multiplicative factor is fixed by demanding the Bekenstein-Hawking law
$s={1 \over 4G}$ at the horizon.  Combining with \wwsax, we deduce that on $\Sigma_c$
\eqn\fdas{{\eta (r_c)\over s(r_c)}={1 \over 4 \pi}}
for any value of $r_c$.\foot{Closely related observations were made in a  different formalism by Iqbal and Liu \IqbalBY.}  We have already seen in our formalism that \fdas\ for the special value $r_c=r_h$ is a universal feature of Rindler space. Now we see that under quite general assumptions, it does not change under RG flow, and so the value \fdas\ will also apply to $r_c=\infty$, in agreement with findings in \refs{\IqbalBY, \BuchelVZ}. So far we have not asked the question: $why$ should radial evolution be isentropic?
A physical answer from the gravity side is that the only entropy associated with a classical solution is the horizon of the black hole. Therefore the total entropy inside any hypersurface outside the horizon should be independent of radius.

This also gives a clue as to when the relation \fdas\ might be violated \MyersYI. When quantum corrections are included on the gravity side, the entropy will generically depend on radius. For example there might be a thermal gas of Hawking radiation surrounding the black hole or
entanglement entropy across $\Sigma_c$. These are suppressed by a factor of $\hbar$ relative to the horizon entropy. We then see no reason to expect that
\fdas\ should survive such quantum corrections.  The universality of \fdas\ is presumably a statement about the classical gravity limit. We note of course that the classical gravity limit is in general not the same as the classical limit of the quantum theory underlying the fluid.

%
%

 \centerline{\bf Acknowledgements}

    We are grateful to S. Cremonini, T. Damour, F. Denef, B. Freivogel, S. Gubser, S. Hartnoll, S. Kachru, H. Liu, R. Loganayagam, A. Maloney, S. Minwalla, P. Petrov, S. Sachdev, L. Susskind, K. Thorne and E. Verlinde for useful discussions over the two years this paper has been incubating.  This work was supported by DOE grant DE-FG0291ER40654 and the Fundamental Laws Initiative at Harvard.
 \listrefs
 \end